\documentclass[prl,showpacs,amssymb,twocolumn,aps]{revtex4}
\usepackage{graphicx,graphics}% Include figure files
\usepackage{epsfig}% Include figure files
\usepackage{dcolumn}% Align table columns on decimal point
\usepackage{bm}% bold math
\usepackage{natbib}
\usepackage{times}
\usepackage{ulem}
\def\kmm#1  {{\bf [KMM:~ #1]~}}
\def\new#1 {{\bf #1 }}
\def\cut#1 {\sout{#1} }
\newcommand{\pks}{PKS~1413+135}

\newcommand{\dal}{\ensuremath{\lsb \Delta \alpha/ \alpha \rsb}}

\newcommand{\dmu}{\ensuremath{\lsb \Delta \mu/\mu \rsb}}
\newcommand{\beq}{\begin{equation}}
\newcommand{\eeq}{\end{equation}}

\newcommand{\lsb}{\left[}
\newcommand{\rsb}{\right]}
\newcommand{\kms}{km~s$^{-1}$}

\begin{document}

\title{Stringent constraints on fundamental constant evolution using conjugate 18\,cm satellite OH
lines}

\author{Nissim Kanekar$^1$, Tapasi Ghosh$^2$, Jayaram N. Chengalur$^1$}
\email{nkanekar@ncra.tifr.res.in}
\affiliation{$^1$National Centre for Radio Astrophysics, Pune 411 007, India; \\
$^2$Arecibo Observatory, Arecibo, PR 00612, USA}

\date{\today}

\begin{abstract}

We have used the Arecibo Telescope to carry out one of the deepest-ever integrations in
radio astronomy, targetting the redshifted conjugate satellite OH~18\,cm lines at 
$z \approx 0.247$ towards PKS\,1413+135. The satellite OH 1720~MHz and 1612~MHz lines are 
respectively in emission and absorption, with exactly the same line shapes due to population 
inversion in the OH ground state levels. Since the 1720 and 1612~MHz line rest frequencies have 
different dependences on the fine structure constant $\alpha$ and the proton-electron 
mass ratio $\mu$, a comparison between their measured redshifts allows one to probe changes in 
$\alpha$ and $\mu$ with cosmological time. In the case of conjugate satellite OH~18\,cm lines, 
the predicted perfect cancellation of the sum of the line optical depths provides a strong 
test for the presence of systematic effects that might limit their use in probing fundamental 
constant evolution. A non-parametric analysis of our new Arecibo data yields 
$\left[\Delta X/X \right] = (+0.97 \pm 1.52) \times 10^{-6}$, where 
$X \equiv \mu \alpha^2$.  Combining this with our earlier results from 
the Arecibo Telescope and the Westerbork Synthesis Radio Telescope, we obtain 
$\left[\Delta X/X \right] = (-1.0 \pm 1.3) \times 10^{-6}$, consistent with no changes 
in the quantity $\mu \alpha^2$ over the last 2.9~Gyr. This is the most 
stringent present constraint on fractional changes in $\mu \alpha^2$ from 
astronomical spectroscopy, and with no evidence for systematic effects.

\end{abstract}
\maketitle

{\it Introduction.}--- Over the last two decades, astronomical spectroscopy of 
high-redshift galaxies has provided the most sensitive probe of changes in fundamental 
``constants'', such as the fine structure constant $\alpha$ and the proton-electron 
mass ratio $\mu \equiv m_p/m_e$, with cosmological time. Such temporal changes in 
low-energy constants like $\alpha$ and $\mu$ are a generic feature of higher-dimensional 
theories aiming to unify general relativity and the standard model of particle 
physics (e.g., \cite{marciano84,damour94}), and are, hence, of much interest. 
The astronomical studies are of particular importance as they allow us to test 
for changes in the constants on Gyr time scales, which are typically inaccessible 
to laboratory studies (e.g., \cite{uzan11}).

The above astronomical techniques are based on comparisons between the measured
redshifts of different spectral lines in high-redshift galaxies, using transitions whose 
rest frequencies have different (and known) dependences on a given constant. At radio 
frequencies, a variety of methods, based on various atomic and molecular lines
\cite{drinkwater98,chengalur03,flambaum07b,jansen11}, have been used to probe temporal 
changes in $\alpha$ and $\mu$. For example, comparisons between inversion and 
rotational transitions in the $z = 0.685$ gravitational lens towards B0218+357 and 
between different methanol (CH$_3$OH) transitions in the $z = 0.886$ lens towards 
B1830$-$21 have yielded the most stringent constraints on changes in $\mu$ from any 
technique, $\dmu < 4 \times 10^{-7}$ \cite{kanekar11,bagdonaite13b,kanekar15}. Comparisons 
between the hydroxyl (OH) 
and H{\sc i} 21\,cm lines in the $z = 0.765$ lens towards PMN~J0134$-$0931 have yielded 
stringent constraints on changes in both $\alpha$ and $\mu$ \cite{kanekar05,kanekar12}. 
And a comparison between the the redshifts of ``conjugate satellite'' hydroxyl (OH) 18\,cm lines 
from the $z = 0.247$ system towards PKS1413+135 has yielded tentative evidence 
(at $\approx 2.6\sigma$ significance) for changes in $\alpha$ and/or $\mu$ with cosmological
time \cite{kanekar10b}.

Amongst the various astronomical methods to probe fundamental constant evolution, techniques 
based on comparisons between multiple spectral lines of a single species (e.g., CH$_3$OH, OH, etc) 
are the least prone to systematic effects. 
An interesting situation arises in the case of the satellite OH~18\,cm lines, which are 
conjugate under certain astrophysical conditions, i.e., have the same shape but opposite sign, with 
one line in emission and the other in absorption, such that the sum of the two optical depths exactly 
cancels out \cite{elitzur92,langevelde95,kanekar04b}. This arises due to population inversion
in the ground state of the OH radical, due to quantum mechanical selection rules (when the 
infrared OH rotational lines that connect the OH ground state to the lower excited states 
are optically thick \cite{elitzur92}). Redshifted conjugate satellite OH lines provide an 
excellent probe of fundamental constant evolution, as the two satellite 
OH line frequencies have different dependences on $\alpha$ and $\mu$ \cite{chengalur03,kanekar04b} 
and the conjugate behaviour guarantees that the lines arise from the same gas. If either $\alpha$ 
or $\mu$ were to change with cosmological time, the satellite line shapes should remain the same but 
the two lines would be offset from each other in velocity space. Conversely, any local systematic 
effects that might give rise to velocity offsets between the lines would also be expected to change the 
line shapes. This implies that the sum of the satellite optical depths would not cancel perfectly 
in the presence of such systematic effects (i.e., the lines would not remain conjugate). 
The cancellation of the sum of the satellite OH~18\,cm optical depths, along with a velocity 
offset between the two lines, is thus a signature of changes in $\alpha$ and/or $\mu$.

% Finally, the two satellite OH lines arise at very nearby 
% frequencies (1612.230825~(15)~MHz and 1720.529887~(15)~MHz; \cite{lev06}), at which the background
% radio source would be expected to have a similar spatial structure.

At present, the sole perfectly conjugate satellite OH~18\,cm system known at cosmological distances is 
the $z = 0.247$ absorber-emitter towards \pks\ \cite{kanekar04b}. Our earlier deep Westerbork 
Synthesis Radio Telescope (WSRT) and Arecibo Telescope observations of this system \cite{kanekar10b}
yielded tentative evidence, at $2.6\sigma$ significance, of a velocity offset between the 
two OH lines, but with the same line shapes, the signature expected from fundamental 
constant evolution. We report here further observations of this system, yielding one of the 
deepest-ever Arecibo Telescope integrations, that allow us to probe changes in 
$X \equiv \mu\alpha^2$ over a lookback time of 2.9~Gyr.

\begin{figure*}
\epsfig{file=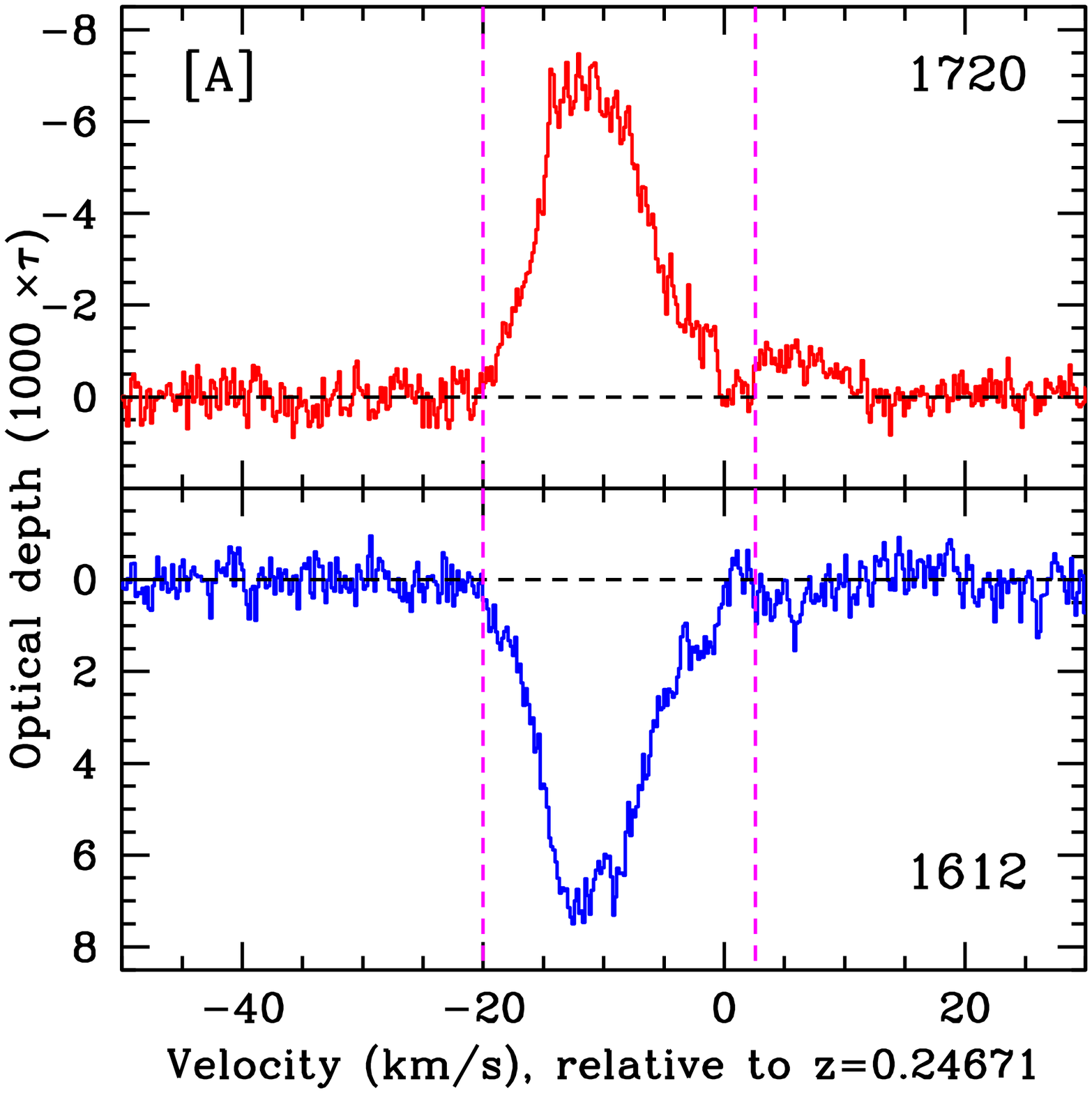,width=3.5in,height=3.5in}
\epsfig{file=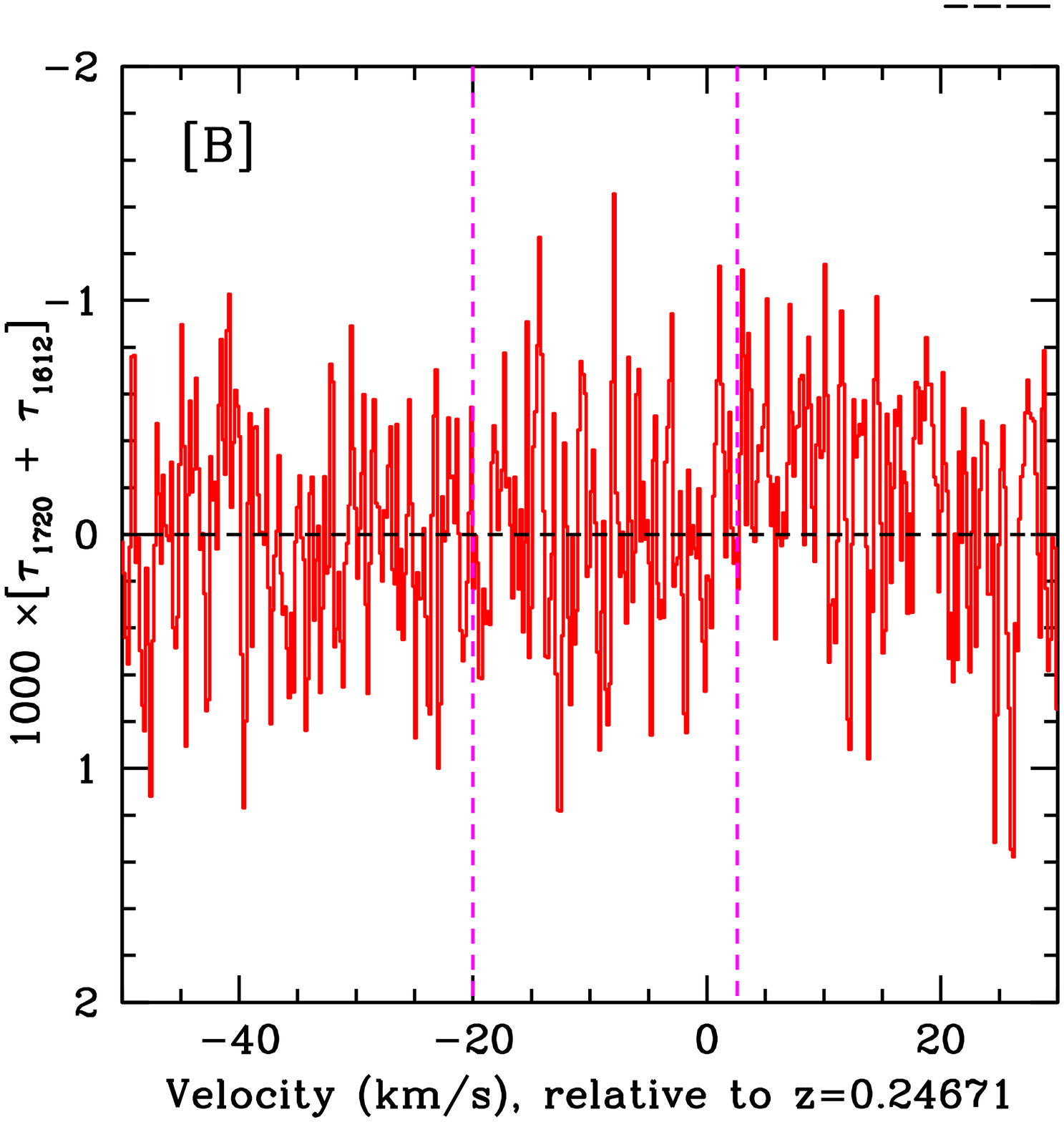,width=3.5in,height=3.5in}
\caption{[A]~(left panel) Arecibo Telescope satellite OH~18\,cm spectra towards \pks, with optical 
        depth ($1000 \times \tau$) plotted against heliocentric velocity, in km/s, relative 
        to $z = 0.24671$. [B]~(right panel) The sum of the 1612 and 1720~MHz optical depth spectra; 
	this is seen to be consistent with noise. In both figures, the dashed vertical lines indicate 
	the velocity range that was used for the cross-correlation analysis.}
\label{fig:fig1}
\end{figure*}

{\it Spectra and results.}--- The Arecibo Telescope was used to carry out a 125-hour integration
on the satellite OH~18\,cm lines of \pks\ between April~2010 and June~2012. The observations 
were carried out in double-position-switched mode \cite{ghosh02}, with five-minute On and Off 
scans on \pks\ followed by two-minute On and Off scans on a nearby bandpass calibrator, PKS~1345+125.
The 1720~MHz and 1612~MHz lines were observed simultaneously on all runs, using the WAPP backends, 
with bandwidths of 1.5625~MHz sub-divided into 4096 channels and centred on the redshifted 
satellite line frequencies. This yielded a velocity resolution of $\approx 0.18$~km/s after 
Hanning smoothing and re-sampling. System temperatures were measured using a noise diode. 

The initial data analysis was carried out in IDL, using standard procedures to produce calibrated 
spectra for the individual double-position-switched scans. Each 5-minute source spectrum was 
then visually inspected for the presence of any systematic effects (e.g., radio frequency interference, 
complex spectral baselines, etc); all spectra showing such effects were excluded from the analysis. 
Each 5-minute spectrum was then subjected to both the Kolmogorov-Smirnov and Anderson-Darling tests,
to test for Gaussianity; spectra failing these tests were also excluded. The above procedure was 
carried out for the two satellite spectra independently; any 5-minute spectrum excluded for
one line was also excluded for the other, to ensure that the final spectra for the two lines 
are based on simultaneous data and, hence, that line variability is not an issue. After all 
data editing (which excised $\lesssim 10$\% of the data), each 5-minute spectrum was shifted to 
the heliocentric frame. The spectra were then converted to optical depth units, and finally 
averaged together for each line. For the 1720~MHz spectra, the weights for the averaging were 
determined from the measured root-mean-square (RMS) noise values; the same weights were used 
when averaging the 1612~MHz spectra, to ensure simultaneity of the final satellite OH~18\,cm spectra.

% We note, in passing, that the shapes of the two satellite OH~18\,cm lines in the new Arecibo spectra 
% are clearly different from those in our earlier WSRT and Arecibo Telescope spectra (from 2005 -- 2009),
% with the peak optical depths of the new spectra lower by $\approx 10$\%. This emphasizes the 
% requirement for simultaneous observations of the satellite OH lines when probing fundamental 
% constant evolution, to ensure that line variability does not affect the results.

The top and bottom panels of Fig.~\ref{fig:fig1}[A] show our final Arecibo Telescope 
satellite OH~18\,cm optical depth spectra, with optical depth plotted versus velocity, in the 
heliocentric frame, in \kms, relative to $z=0.24671$. The spectra have RMS optical depth noise 
values of $3.6 \times 10^{-4}$ (1612~MHz) and $3.1 \times 10^{-4}$ (1720~MHz), per 
$\approx 0.18$~\kms\ channel. The Anderson-Darling test finds that the off-line regions 
of both spectra are consistent with arising from Gaussian noise;
there is no evidence for systematic structure in either spectral baseline. Fig.~\ref{fig:fig1}[B]
shows the sum of the 1612 and 1720~MHz optical depth profiles, again plotted against heliocentric 
velocity, in \kms, relative to $z=0.24671$. The RMS noise on the summed spectrum is 
$5.1 \times 10^{-4}$ per $0.18$~km/s channel, consistent with the RMS noise values on 
the individual satellite spectra. The summed spectrum shows no evidence for non-Gaussian structure: 
both the Kolmogorov-Smirnov rank-1 test and Anderson-Darling test find that the summed 
spectrum is consistent with being drawn from a normal distribution. We thus find that the 
satellite OH~18\,cm lines remain conjugate at the signal-to-noise ratio of our new Arecibo 
Telescope observations.

Most astronomical techniques probing fundamental constant evolution are based on the 
modelling of the line profiles as the sum of Gaussians or Voigt profiles. Multi-component 
fits are used to estimate the redshifts of the individual spectral components (e.g., 
\cite{murphy03,kanekar10,kanekar12,murphy16}). For complex profiles, this process of fitting multiple 
spectral components can itself affect the results (e.g., due to under-fitting or 
over-fitting the line profile). However, in the case of conjugate satellite OH~18\,cm lines,
the maser mechanism that gives rise to the conjugate behaviour guarantees that the lines have exactly 
the same shape. This implies that a non-parametric technique, based on cross-correlation of the 
two spectra, can be used to measure the redshift difference between the lines, and, hence, to probe 
changes in $\alpha$ and $\mu$ \cite{kanekar10b}. Specifically, the velocity offset from zero of the 
peak of the cross-correlation of the two OH lines directly yields the redshift difference between 
the lines. The fact that the cross-correlation technique is non-parametric and, hence, not susceptible
to errors regarding the decomposition of a line into multiple spectral components is an important
advantage of conjugate satellite OH lines over other approaches (e.g., the many-multiplet method; 
\cite{dzuba99,murphy04}) in probing fundamental constant evolution.

We used the velocity range $-20$~\kms\ to $+2.6$~\kms\ (indicated by the dashed vertical lines 
in Fig.~\ref{fig:fig1}[B]), enclosing the strongest spectral line feature, for the 
cross-correlation; this was done in order to maximize the signal-to-noise ratio. The offset 
of the peak of the cross-correlation from zero velocity was measured via a Gaussian fit. 
Very similar results were obtained on using other functional forms (e.g., a parabolic form) 
for the cross-correlation as well as other non-parametric methods (e.g., the sliding distance method 
\cite{levshakov12}). The RMS noise on the cross-correlation was obtained via a Monte Carlo approach, 
in which we cross-correlated 10$^4$ pairs of simulated satellite OH~18\,cm spectra. The simulated 
spectra were obtained by adding 
Gaussian random noise (characterized by the RMS noise values on the observed spectra) to the best 
4-component Gaussian fits to the 1720~MHz and 1612~MHz spectra. Note that the Gaussian fits were only 
used to obtain templates for the two spectra, and do not affect the results in any way. We find that 
the cross-correlation of the two satellite OH~18\,cm lines peaks at a velocity offset of 
$\Delta V_{\rm new} = (+35.0 \pm 56.5)$~m/s (all quoted errors are at $1\sigma$ significance), with 
the 1720~MHz line at a lower 
velocity (i.e.  at a lower redshift). Our present Arecibo observations thus find no evidence of a 
statistically significant velocity offset between the 1612~MHz and 1720~MHz spectra. Using equation~(13) of 
\citet{chengalur03} then yields $\lsb \Delta X/X \rsb = (+0.97 \pm 1.52) \times 10^{-6}$, 
where $X \equiv \mu \alpha^2$.

Our earlier WSRT and Arecibo studies of the OH~18\,cm lines from \pks, carried out between
2006 and 2008, yielded a net velocity offset of $\Delta V_{\rm old} = (-230 \pm 90)$~m/s 
between the two satellite lines \citep{kanekar10b}, yielding $\lsb \Delta X/X \rsb = 
(-6.3 \pm 2.5) \times 10^{-6}$. Combining the old and the new results for $\lsb \Delta X/X \rsb$, 
with appropriate weights based on the RMS noise values, our final result is 
$\lsb \Delta X/X \rsb = (-1.0 \pm 1.3) \times 10^{-6}$. We thus find no statistically significant
evidence for changes in $X \equiv \mu \alpha^2$ over a lookback time of 2.9~Gyr.

% We also carried out a similar cross-correlation analysis on the strong conjugate satellite 
% OH lines from the nearby extra-galactic source Cen.A \cite{langevelde95}, to test whether 
% velocity offsets between the satellite lines are seen in local conjugate systems. The analysed 
% OH spectra towards this source were kindly provided to us by Huib van Langevelde. The peak of 
% the cross-correlation for the Cen.A spectra was found to be at 
% $\Delta V = (+0.05 \pm 0.11)$~km/s, consistent with no velocity offset between the lines. 
% This demonstrates that the conjugate satellite technique provides the expected null result 
% for at least one local system, and at a sensitivity comparable to that of the datasets towards 
% \pks.

{\it Systematic effects.}--- Systematic effects in the conjugate satellite OH method that 
might contribute to increased errors, over and above those determined from the 
cross-correlation analysis, are discussed in detail in \cite{kanekar10b}. The systematics
fall in two broad categories, those arising from observational issues [e.g., doppler tracking, 
frequency calibration, errors in the laboratory frequencies, radio frequency interference (RFI), etc]
and astronomical issues (e.g., different intrinsic shapes of the two satellite OH lines, 
interloping lines from other transitions, etc). Errors arising from the above observational 
systematics are small compared to our measurement errors. Specifically, doppler corrections for 
Earth motion were carried out offline, using a model of Earth motion accurate to $< 15$~m/s, 
a factor of four smaller than our measurement error. The Arecibo Telescope frequency scale 
is set by the accuracy of masers and local oscillators (1~Hz), more than two orders of magnitude 
smaller 
than the measurement error. The satellite OH line frequencies have been measured in the laboratory 
with an accuracy of $\approx 15$~Hz \cite{lev06}, more than an order of magnitude smaller than 
our measurement error. Finally, detailed statistical tests for non-Gaussian behaviour that might 
arise from RFI were carried out on both the individual 5-m spectra and the final averaged 
satellite OH spectra, and only spectra that passed these stringent tests were retained in 
the analysis. The fact that the satellite OH lines remain conjugate, with the sum of the two 
optical depth spectra consistent with random noise, indicates that RFI is not an important 
issue for these data. 

Considering the second category, astronomical effects, \citet{kanekar10b} note that there is 
no possibility of line interlopers from other spectral transitions, either from the Milky 
Way or galaxies at different redshifts along the line of sight to \pks, or from other 
molecular or atomic lines from \pks\ itself. And, as in the case of RFI, the strongest 
argument against the possibility of different intrinsic structure in the two satellite 
OH line profiles (or line interlopers) is the fact that the sum of the optical depth spectra 
is consistent with Gaussian noise, exactly as predicted by the maser mechanism for conjugate behaviour. 
This provides a stringent test for the use of the conjugate satellite OH lines to probe fundamental 
constant evolution: the satellite OH lines of \pks\ pass this test at our current sensitivity. 
Finally, an important test of the use of any technique to probe temporal evolution 
in the fundamental constants is that the same technique yield a null result in the local 
Universe. For the conjugate satellite OH technique, the expected null result was indeed 
obtained by \citet{kanekar10b} for the nearby conjugate satellite OH system in Cen.A 
(whose OH lines were observed with the Australia Telescope Compact Array; \cite{langevelde95}), at 
a sensitivity similar to that of the present Arecibo Telescope spectra. Overall, we find no 
evidence that our result might be affected by systematic effects, related to either 
observational or astronomical issues.

{\it Discussion}--- A wide variety of techniques, at optical and radio wavelengths, 
have been used to probe the possibility of temporal changes in $\alpha$ and $\mu$, 
or combinations of these quantities. In the optical regime, using echelle spectrographs 
on 10m-class optical telescopes, the many-multiplet method 
\cite{dzuba99}, based on rest-frame ultraviolet spectral lines,  has yielded the 
highest sensitivity to changes in $\alpha$ out to relatively high redshifts, $z \approx 3$ (e.g.,  
\cite{murphy04,webb11,molaro13,evans14,murphy16,kotus17}). 
% \cite{murphy04,levshakov06,webb11,molaro13,evans14,wilczynska15,murphy16,kotus17}). 
The most sensitive results have {\it statistical errors} of $\dal \approx (1-2) \times 10^{-6}$, 
either based on individual systems (e.g.,  \cite{levshakov06,kotus17}) or large 
absorption samples (e.g., \cite{murphy04,webb11}). Conversely, redshifted ultraviolet 
ro-vibrational molecular hydrogen (H$_2$) lines have yielded the highest sensitivity to changes 
in $\mu$, again out to $z \approx 3$ (e.g., \cite{reinhold06,vanweerdenburg11,rahmani13,dapra17}). 
% (e.g. \cite{reinhold06,malec10,king11,vanweerdenburg11,wendt11,rahmani13,bagdonaite15,dapra17}). 
The most sensitive of the results here have yielded statistical errors of 
$\dmu \approx 2 \times 10^{-6}$ (e.g., \cite{ubachs16}).

Unfortunately, while the above optical results have low {\it statistical} errors, it has 
recently become clear that most of these studies are afflicted by {\it systematic} errors 
(e.g., \cite{griest10,whitmore10,whitmore15}). The problem here has to do with the wavelength 
calibration of the optical echelle spectrographs, mostly the Keck Telescope High Resolution Echelle 
Spectrograph (HiRES) and the Very Large Telescope (VLT) UltraViolet Echelle Spectrograph (UVES), 
that were used for the optical observations. As noted in \cite{whitmore15,murphy16}, 
{\it all} $\dal$ (and $\dmu$) results derived from Keck-HIRES and VLT-UVES spectroscopy 
until 2014 are likely to be affected by systematic errors due to long-range distortions in 
the wavelength calibration. These distortions are still not understood and it is not possible
in most cases to retrospectively correct the earlier spectra \cite{murphy16}. Recently, the 
many-multiplet method has been used with ``super-calibration'' techniques or comparisons between 
very nearby lines to reduce the effects of the above long-range distortions (e.g., 
\cite{evans14,murphy16,murphy17}). For example, a comparison between Zn{\sc ii} and Cr{\sc ii} 
lines in Keck-HIRES and VLT-UVES spectra of nine absorbers yielded $\dal = [+1.15 \pm 1.67 (stat.) 
\pm 0.87 (sys.)] \times 10^{-6}$ \cite{murphy16}. In another study, the long-range distortions in 
multiple VLT-UVES spectra of a single bright quasar (taken over a ten-year period) were corrected 
using high-accuracy spectra from the HARPS spectrograph. The many-multiplet method was then 
applied to these spectra, to obtain a high sensitivity to changes in $\alpha$, $\dal = 
[-1.42 \pm 0.55 (stat.) \pm 0.65 (syst.)] \times 10^{-6}$ \cite{kotus17}. 

At radio wavelengths, the CH$_3$OH and NH$_3$ techniques have yielded stringent constraints 
on changes in $\mu$, with $\dmu \lesssim 4 \times 10^{-7}$ from individual absorbers at 
intermediate redshifts, $z \approx 0.685$ (NH$_3$; \cite{kanekar11}) and $z \approx 0.886$ 
(CH$_3$OH; \cite{kanekar15}). The CH$_3$OH technique is perhaps the most interesting 
of these radio methods as it yields both high sensitivity and a good control of systematic effects, 
as thermally-excited and optically-thin spectral transitions of a single molecule are used in 
the analysis, and one can test that the different lines have the expected ratios in thermal 
equilibrium \cite{kanekar15}. 

In the case of the conjugate satellite OH lines, the technique is sensitive to changes in 
$X \equiv \mu \alpha^2$, and does not provide constraints 
on changes in the individual constants, without further assumptions. Our result implies 
$2.0\dal + \dmu = (-1.0 \pm 1.3) \times 10^{-6}$ over $0 < z < 0.247$. This implies $1\sigma$ 
sensitivities of $\dal = 0.65 \times 10^{-6}$ (if we assume $\dmu = 0$) and 
$\dmu = 1.3 \times 10^{-6}$ (if we assume that $\dal = 0$).
The crucial advantage of this method is that it allows one to directly test whether it 
can be applied at all, via the prediction that the two satellite OH lines must 
have the same shapes, with opposite signs. Like the CH$_3$OH method discussed above, 
this technique allows one to measure 
changes in the constants from a single space-time location, without the need to average 
over multiple absorbers (as would be required in most other techniques to overcome possible 
local velocity offsets between the gas clouds giving rise to the different lines). These 
two techniques are, hence, especially interesting to probe the possibility of space-time 
variation in $\alpha$ and $\mu$.

Our results are based on one of the deepest-ever observations with the Arecibo Telescope, 
which has the largest collecting area and sensitivity of today's radio telescopes. 
However, new radio telescopes [e.g., the Five-Hundred-Metre Aperture Spherical Telescope (FAST) 
and the Square Kilometre Array (SKA)] are now being built or planned that will have even higher 
sensitivity than the Arecibo Telescope; these will allow both higher sensitivity on known 
conjugate satellite OH 18\,cm systems like \pks, and searches for new conjugate satellite 
systems at high redshifts. Modern high-frequency radio telescopes like the Very Large Array 
should allow an improvement in sensitivity to fractional changes in $\mu$ by more than an 
order of magnitude, using the NH$_3$ and CH$_3$OH lines. Finally, the combination of high 
sensitivity and new wavelength calibration schemes on next-generation large optical 
telescopes (e.g., the Thirty Meter Telescope, the Giant Magellan Telescope and the European
Extremely Large Telescope) should also allow improvements in the sensitivity to fractional
changes in $\alpha$ and $\mu$ by $1-2$ orders of magnitude via the many-multiplet
method and ro-vibrational H$_2$ lines.

In summary, we have carried out an ultra-deep Arecibo Telescope observation 
of the OH~18\,cm satellite lines from \pks\ at $z \approx 0.247$. We find that the 
satellite OH lines are conjugate within our measurement errors, with the 1720~MHz 
line in emission, the 1612~MHz line in absorption, and the sum of the two optical 
depth spectra consistent with Gaussian noise. We used a non-parametric technique, based on 
cross-correlation, to test for a velocity offset between the two OH lines, that might 
arise due to changes in $\alpha$ and/or $\mu$. The cross-correlation analysis finds that
the velocity offset between the lines is $\Delta V_{\rm new} = (+35.0 \pm 56.5)$~m~s$^{-1}$, 
consistent with the null hypothesis of zero velocity offset between the OH 1612~MHz and 
1720~MHz lines. This implies $\lsb \Delta X/X \rsb = (+0.97 \pm 1.52) \times 10^{-6}$, 
where $X \equiv \mu \alpha^2$. Combining this with the results from our earlier 
Arecibo/WSRT analysis yields our final result, $\lsb \Delta X/X \rsb = (-1.0 \pm 1.3) 
\times 10^{-6}$ over $0 < z < 0.247$, consistent with no change in $\mu \alpha^2$ 
over a lookback time of $\approx 2.9$~Gyr.

\begin{acknowledgments}
We thank Chris Salter for much help with the Arecibo observations, and two anonymous referees
for useful comments on an earlier version of this paper. NK acknowledges support from a 
Swarnajayanti Fellowship of the Department of Science and Technology (DST/SJF/PSA-01/2012-2013). 
The Arecibo Observatory is operated by SRI International under a cooperative agreement with the 
National Science Foundation (AST-1100968), and in alliance with Ana G. M{\'e}ndez-Universidad 
Metropolitana, and the Universities Space Research Association. 

\end{acknowledgments}

\end{document}